\documentclass[10pt,a4paper,twocolumn]{article}
\usepackage{graphicx}
\usepackage{authblk}% do formatowania afiliacji
\usepackage[english]{babel}
\usepackage{amsmath}
\usepackage{amsfonts}
\usepackage{hyperref}
\usepackage{enumitem}%allows to not have an indent
\title{Cellular Automaton-Like Model of Arising Physical-Like Properties}
\author{M.~Pietrow\footnote{e-mail: mrk@kft.umcs.lublin.pl}}
\affil[]{Institute of Physics, M. Curie-Sk{\l}odowska University, ul.~Pl.~M.~Curie-Sk{\l}odowskiej~1,\\20-031 Lublin, Poland}
\usepackage{draftwatermark}
\SetWatermarkScale{7}
\SetWatermarkText{paper draft}
\SetWatermarkLightness{0.95}
\begin{document}
\maketitle
\begin{figure}
\begin{center}
\includegraphics[width=0.5 \textwidth]{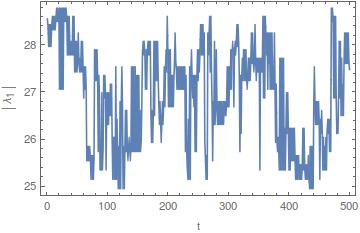}
\end{center}
\caption{Absolute values of one of non-zero eigenvalues for a ten-dimensional $mrel$ during 500 steps of evolution consisting in random swaps of columns and rows (but not governed by the~(\ref{eq:Evolution}) rule). The time-line resembles one-dimensional random walk.}
\label{fig:Evolution6_FlipsRandomLambda}
\end{figure}

\paragraph{Abstract}
A simple relation of the order of $n$ abstract objects generates an $n-2$ dimensional basis of three dimensional vectors. A cellular automaton-like model of evolution of this system is postulated. During this evolution, some quantities stabilise with time and form a discrete spectrum of values. The presented model may have some general aspects in common with a cellular automaton representation of a quantum system.
\paragraph{Introduction:}
Cellular automata (CA) are used to describe the behaviour of systems with a wide range of complexity from physics to biology~\cite{Wolfram02}. Mainly, the description is functional, but not a structural one (CA rules allow description of some aspects of a system at a high structural level without references to the rules from the deeper level of subsystems).\\
The aim of this presentation is opposite to some extent. One does not require here a compatibility of an introduced model with any special real system. Instead, it was assumed that there exists a set of some abstract objects and a general quantity, an order, characterising each member of this set. Based on this, a matrix was constructed which keeps relations between these objects. The properties of this matrix were examined and some arising similarities to physical properties were brought into focus. Furthermore, the presented model keeps the compatibility with CA ideas to some extent and tries to adhere to a description of a set of physical objects from the structural point of view.\\
Although most of the ideas here are postulated but not derived from something deeper, it would be promising to consider the Author's idea of a simple relation between some elementary objects and an introductory model of how physical properties arise from it.
\paragraph{Relation matrix ($mrel$):}
A fundamental feature of a system of basic objects (thought here as abstract entities, not physical ones; called here \emph{elementary objects}) is the relation between them. One of the simplest relations seems to be an 'order' of these objects. For example, for three objects there are 3$!$ of their possible arrangements.\\
Now, consider a set of $n$ identical elementary objects. Let us define an $n\times n$ matrix $mrel_{i,j}$ (\emph{relation matrix}), which describes the distance (in the meaning of this order) of the $i$-th relative to the $j$-th object. For example, $mrel_{1,2}$=-1 because the $1^{\textrm{st}}$ object is one step before the $2^{\textrm{nd}}$ one (it proceeds object $2$). The $mrel$ for three particles of the order $\{1,2,3\}$ is
\begin{equation}
\left(
\begin{array}{ccc}
 0 & -1 & -2 \\
 1 & 0 & -1 \\
 2 & 1 & 0
\end{array}
\right),
\end{equation}
whereas for the $\{1,3,2\}$ order we have
\begin{equation}
\left(
\begin{array}{ccc}
 0 & -2 & -1 \\
 2 & 0 & 1 \\
 1 & -1 & 0
\end{array}
\right).
\end{equation}
\paragraph{The Eigensystem of the $mrel$:}
$Mrel$s have interesting properties. Consider a $5\times 5$ $mrel$ for an arrangement $\{1,2,3,4,5\}$ as an example. Its eigenvalues are
\begin{equation}
\{\lambda_1=5i\sqrt{2},\lambda_1^{\ast},0,0,0\},
\end{equation}
whereas the corresponding eigenvectors are
\begin{eqnarray}
v_1= & \{\frac{1}{3}(-1+2i\sqrt{2}),\frac{i}{\sqrt{2}},\frac{1}{3}(1+i\sqrt{2}),\frac{1}{6}(4+i\sqrt{2}),1\},\nonumber\\
v_2= & v_1^{\ast},\\
v_3= & \{3,-4,0,0,1\}, v_4=\{2,-3,0,1,0\}, v_5=\{1,-2,1,0,0\},\nonumber
\end{eqnarray}
where $^{\ast}$ denotes a complex conjugation.\\
The following is an interesting general rule for $mrel$ (no matter what its $n$ dimension is). Its $n-2$ eigenvalues are equal to $0$, whereas the respective eigenvectors \emph{always have the non-zero values in 3 dimensions only}. These vectors span the $n-2$ space, $physV$.\\
The $physV$ seems to be a promising representation of $n-2$ physical objects in a common physical space, i.e. these $n-2$ eigenvectors can describe the basic objects in a three-dimensional sub-spaces of a common space. Let us call these vectors with the zero eigenvalues the \emph{physical vectors}.
\paragraph{Other properties of $mrel$:}
Some other interesting properties of $mrel$ are listed below.
\begin{enumerate}[leftmargin=*]
\item Permutation of related elementary objects does not change the eigenvalues of $mrel$, whereas the eigenvectors do change.
%\item Flips of columns or rows of a $mrel$ change the physical vectors inside a set $\{-v,0,v\}$.
\item The physical vectors are independent of $a_0$ in the case of generalisation of the relation definition in $mrel$ as
\begin{multline}
\cdots, -1\rightarrow a_0-1, 0\rightarrow a_0, 1\rightarrow a_0+1,\\
2\rightarrow a_0+2, etc..
\end{multline}
\item Any $mrel$'s sub-matrix of dimension $n'$ has $n'-2$ physical vectors.
%\begin{enumerate}[leftmargin=*]
\item $2\times 2$ $mrel$s (for the arrangements $\{1,2\}$ and $\{2,1\}$) have no physical vectors. Normalised eigenvectors of these matrices resemble spin vectors for a spin-$\frac{1}{2}$ particle
\begin{equation}
\{-\frac{i}{\sqrt{2}},\frac{1}{\sqrt{2}}\},\quad \{\frac{i}{\sqrt{2}},\frac{1}{\sqrt{2}}\},
\end{equation}
whereas these $mrel$s are proportional to one of the Pauli matrices, $\sigma_2$: $mrel(\{1,2\})=i\sigma_2$ and $mrel(\{2,1\})=-i\sigma_2$.\\
For three elementary objects in the relation, there is one physical vector as an eigenvector\footnote{When rearrangement of the elementary objects takes place the components of this vector
\begin{equation}
\frac{1}{\sqrt{6}}\ \{1,-2,1\}
\end{equation}
interchange.
}. In this case, all $2\times 2$ sub-matrices of all three-dimensional $mrel$s generated from permutations of the order $\{1,2,3\}$ give the eigenvalues from the set $\{-2i,-i,1,2\}$ and these sub-matrices are a simple combination of the Pauli matrices. For $dim(mrel)>3$ (two or more physical vectors present) the expansion into the Pauli matrices becomes less trivial.\\
To generalise, the $2 \times 2$ (sub)-$mrel$s seem to be promising operators for spin description.\\
The time evolution of such a system is postulated below.
%\end{enumerate}
\item $Mrel$s are antihermitian (antisymmetric). Some sets of $mrel$s form a linearly independent set (for example, a subset of three $mrel$s generated by permutations of elementary objects). According to the general theory \cite{Antihermitian}, they form a Lie algebra of generators related to some unitary matrices. This suggests a possibility of description of quantum-like evolution~\cite{Greiner94} by these matrices.
\item\label{AlaMaKota} Another scheme of a time evolution (called a \emph{second kind}) of a system described by $mrel$ could be suggested by the case of 2-dim $mrel$s which have been linked with a spin. Each of the Pauli matrices can be derived from one of them by some elementary operations known from linear algebra (two lines\footnote{rows or columns, optionally} switching, a line multiplication by a number). Thus, the evolution of $mrel$ in a general sense could be identified with elementary operations. In general, swapping lines is not equivalent to permutations of the elementary objects.\\
In the simplest case, one may consider a $mrel$ at each step where some two lines could be randomly swapped. However, a more complicated algorithm could be used as a current $mrel$ generator. A new $mrel$ could be considered as a product of up-to-now $mrel$s that could change additionally at some steps by swaps of lines.\\
On the other hand, continuing the idea of relations, for a system of three elementary objects as an example, their states $A$, $B$, $C$ are influenced by each state from all these objects in the set. Thus, it could be written
\begin{align}
A\ =\ & m_{1,1}\ A+m_{1,2}\ B+m_{1,3}\ C,\nonumber\\
B\ =\ & m_{2,1}\ A+m_{2,2}\ B+m_{2,3}\ C,\label{eq:IntoSelf}\\
C\ =\ & m_{3,1}\ A+m_{3,2}\ B+m_{3,3}\ C.\nonumber
\end{align}
The matrix $m_{ij}$ here could be identified with $mrel$.\\
%Additionally, by the analogy to a quantum mechanical we have
%\begin{equation}
%\psi(t+dt)\simeq (1-i H dt/\hbar)\ \psi(t).
%\end{equation}
More generally, for a set of consecutive steps $t$, eq.~(\ref{eq:IntoSelf}) gives
\begin{equation}
\begin{bmatrix}
A\\
B\\
C
\end{bmatrix}=
(mrel)^t\times
\begin{bmatrix}
A\\
B\\
C
\end{bmatrix}.
\label{eq:Evolution}
\end{equation}
The equation above is, in fact, a requirement to find a vector $[A,B,C]^T$ which is unchanged by a projection by the $mrel^t$ operator. Vectors which are the solution of~(\ref{eq:Evolution}) have interesting properties.\\
As an example, consider the $mrel$ for three elementary objects. Calculate $B$ and $C$ as a function of time (because the rank of any $mrel$ is 2, $B$ and $C$ are $A$--dependent here). These functions are shown in fig.~\ref{fig:Evolution}.
\begin{figure}
\begin{center}
\includegraphics[width=0.5 \textwidth]{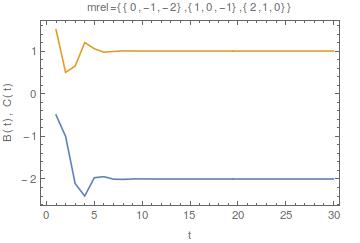}
\end{center}
\caption{Solution of eq.~(\ref{eq:Evolution}) for $3\times 3$ $mrel$ for some $t$.}
\label{fig:Evolution}
\end{figure}
%Ewolution1_MrelPowers.nb
Additionally, the physical vectors \emph{do not} change with steps, whereas the rest of the eigenvectors set oscillate within some set of values.\\
The non-zero eigenvalues of $mrel^t$ rise logarithmically with steps when the system evolves without swaps in between inside the matrix--fig.~\ref{fig:LogEigenvalues}.
\begin{figure}
\begin{center}
\includegraphics[width=0.5 \textwidth]{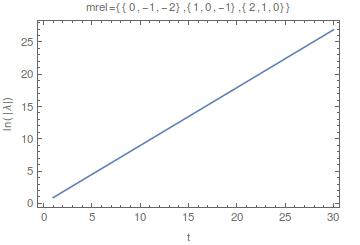}
\end{center}
\caption{Logarithm of an absolute of a non-zero eigenvalue for $3\times 3$ mrel with steps.}
\label{fig:LogEigenvalues}
\end{figure}
However, when the swaps of lines take place, the non-zero eigenvalue rises much faster that logarithmically.\\
%%%%%%%%%% SWAPS %%%%%%%%%%%%%%%%%%%%%%%%%%%%%%%%%%%%%%
One may consider the evolution complicated one step more. If one makes some swaps of lines and \emph{then} solves eq.~(\ref{eq:Evolution}), the result for $B$ and $C$ will approximate asymptotically some value -- fig.~\ref{fig:EvolutionFlipped}.
\begin{figure}
\begin{center}
\includegraphics[width=0.5 \textwidth]{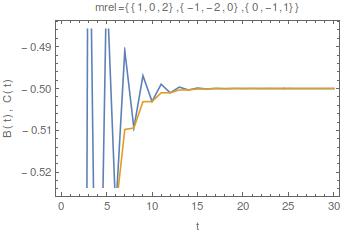}
\end{center}
\caption{Values of $B(t)$, $C(t)$ for a $3\times 3$ $mrel$ in which some swaps of columns and rows precede the evolution.}
\label{fig:EvolutionFlipped}
\end{figure}
%Ewolution1_MrelPowers.nb
\\
On the other hand, if one makes a swap of matrix lines between some steps of evolution and solves eq.~(\ref{eq:Evolution}) after each step then one observes switches to some other value for some time (fig.~\ref{fig:EvolutionFlippedDiscreteLevels}). The interesting feature of this evolution is that the spectrum of values is discrete (they form a multiplet).
\begin{figure}
\begin{center}
\includegraphics[width=0.5 \textwidth]{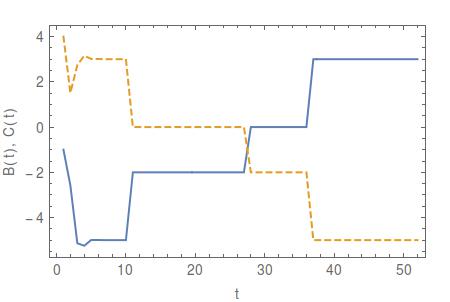}
\end{center}
\caption{Values of $B(t)$, $C(t)$ for a four-dimensional $mrel$ where eleven random swaps of columns and rows were made during the evolution.}
\label{fig:EvolutionFlippedDiscreteLevels}
\end{figure}
%Ewolution2_MrelPowers.nb
%
Generally, the changes of values do not coincide with the moment of the swap of the matrix lines.\\
A discrete spectrum of $B(t)$ and $C(t)$ is also obtained when one calculates it for any sub-matrix of a larger $mrel$ under evolution.\\
The evolution of a second kind erases the anti-symmetricity of a $mrel$ and thus it is a considerably different scheme. However, the antisymmetricity returns after some swaps.
\item If the evolution consists in swapping lines, the number of $n-2$ physical vectors does not change. Also, if one considers the evolution (\ref{eq:Evolution}) with $mrel^t$, the number of the physical vectors remains constant.
\begin{figure}
\begin{center}
\includegraphics[width=0.4 \textwidth]{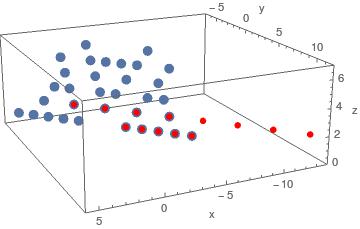}
\end{center}
\caption{A part of the network of points (blue) generated by any $\zeta_i$ set (see description in the text) for permutations of $n$=7 elementary objects. This network could be increased by additional points obtained for larger $n$. The red points are a set of coordinates for physical vectors for a $mrel$ generated by 1000 random swaps of lines of an initial $mrel$ for randomly given permutation of $n$=15 elementary objects.}
\label{fig:physVnetwork}
\end{figure}
%Mrel_CzySwapSpowodujeWypadnieciePhysVPozaPlaszczyzne.nb
\\
It is interesting to consider physical vectors relating to $mrel$s representing all permutations of $n$ elementary objects. These vectors form sets with non-zero values at different three of $n$ positions: $\zeta_1: \{[x, y, z, 0, ...]\}$, $\zeta_2: \{[x, y, 0, z, 0, ...]\}$, $\zeta_3: \{[x, y, 0, 0, z, 0, ...]\}$, etc.. Each $\zeta_i$ points the same network of points located at a plane $x+y+z=0$ (blue points in fig.~\ref{fig:physVnetwork}; any length of vectors are possible). The number of points increases with $n$ (all points generated by smaller set of $n$ elementary objects are generated by a larger one, too). Furthermore, any swaps of $mrel$'s lines produce physical vectors which are a subset of the network given by permutations of elementary objects (e.g.: red points in fig.~\ref{fig:physVnetwork}). Moreover, a multiplication of $mrel$ mentioned in the eq.~(\ref{eq:Evolution}) does not give an additional points but those generated by permutations. Permutations and $mrel$ powering (no matter what is done first) give the points from the regular structure whose an initial part was depicted in fig.~\ref{fig:physVnetwork}.\\
In fact, any length of the eigenvectors of $mrel$s are possible. If one limits to normalized vectors only the set of possible points form a part of a circle centred at $(0, 0, 0)$ with radius 1 and normal vector pointing in the direction of $[1, 1, 1]$ (blue points in fig.~\ref{fig:physV_norm}).
\begin{figure}
\begin{center}
\includegraphics[width=0.4 \textwidth]{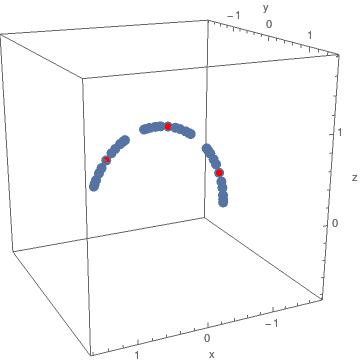}
\end{center}
\caption{The blue points show coordinates of normalised physical vectors during the evolution consisting in $mrel$ lines swapping and powering it randomly. The red points show only states which can be occupied when the $n=3$ problem is considered. These points are vertices of an equilateral triangle.}
\label{fig:physV_norm}
\end{figure}
\\
Let us follow the position of the points described by $\zeta_1$ at each step of the evolution consisting on random swapping lines or powering the matrix. If $n>3$ then the position of the point could change randomly from step to step at a semi-circle of possible points. However, if $n=3$ (there is only one physical vector) only jumps between the points given in red in fig.~\ref{fig:physV_norm} are possible.\\
%On the other hand, if one takes into account all $\zeta_i$ for a $mrel$ at given step of the evolution their positions (placing their non-zero $z$ as it is for $\zeta_1$) cover the whole semi-circle.\\
To generalise, the physical vectors point a net of places in a three-dimensional sub-space for each of $n-2$ objects. The structure of the network (positions of allowed points) is the same for each of these physical vector. According to this, each $\zeta_i$ has its own ('internal') net of possible states. Although each physical vector is represented in its own subspace, from this model, the coordinates of each possible point obey the equation $x+y+z'=0$, where $x$- and $y$-coordinates can be regarded as common ones whereas the $z'$-coordinate is set individually for each vector.\\
The further Author's work will be devoted to check if the jumps through the network (for the one particle case, in particular) could describe a space-time motion of elementary objects in some way.
\end{enumerate}
The evolution described in point \ref{AlaMaKota} above resembles rules obeyed by the CA~\cite{Wolfram02} in general. Its algorithm is an application of a simple rule~(\ref{eq:Evolution}) at each step (however, when swaps of matrix lines take place, randomness of choice as a generalisation of CA rules is added). The equivalence of cells in CA would be matrix elements (or lines) here. Each matrix element changes by application of a rule that requires other elements (but not neighbouring ones only). Additionally, in both cases, the $mrel$ evolution and the CA, some values can form a complex pattern of changes in 'time'. Such behaviour is maintained by non-zero eigenvalues of $mrel$s (fig.~\ref{fig:Evolution6_FlipsRandomLambda}).
%Evolution6_LinesSwap.nb
%
%
%%%%%%%%%%%%%%%%%%%%%%%%%%%%%%%%%%%%%%%%%%%%%%%%%%%%%%
\paragraph{Conclusions:}
This paper presents a collection of statements and hypotheses concerning a relation between basic physical properties, e.g. a number of dimensions of space containing physical objects or an evolution process, and relation matrix properties for which some characteristics have been investigated. From a point of view of the model presented above, the $mrel$ resembles operators in quantum mechanics. Possibly, a permutation group would help to find a link. An interesting consequence would be that the spin-like vector may originate from two-dimensional $mrel$ eigensystems which differ in dimensionality only from three dimensional physical vectors originating from larger $mrel$s.\\
The statements do not form a consistent view of linked concepts but the Author's hope is that the interesting properties of $mrel$s do reveal a structure resembling CA with quantum-like properties and could be developed for a useful description of physical many-body systems.
\bibliographystyle{ieeetr}
\bibliography{references}
\rightline{$\Box$}
\end{document}